\documentclass[
    ,final            
  ]
  {aipproc}

\layoutstyle{6x9}

\begin{document}

\title{The scalar $\kappa$ from $D^+ \to K^-\pi^+\pi^+$:\\ Further Studies}

\author{Carla G\"obel}{
  address={Centro Brasileiro de Pesquisas F\'\i sicas, Rua Dr. Xavier Sigaud 150, Urca \\
  22290-180, Rio de Janeiro, Brasil \\ 
  ~ \\
  Representing the Fermilab E791 Collaboration}
}

\begin{abstract}
We briefly review the recent results obtained by Fermilab experiment E791 on the Dalitz
plot analysis of the decay $D^+\to K^-\pi^+\pi^+$, where indication for a light
$K\pi$ scalar resonance, the $\kappa$, was found. We also present preliminary studies providing 
further information on the phase behavior of the scalar components at low mass, supporting 
the previous indication for the $\kappa$.
\end{abstract}

\maketitle

\section{INTRODUCTION}

Nowadays, with the rising statistics of charm samples, the decays of charm mesons
can be seen as a new source for the study of light meson spectroscopy, complementary to that
from scattering experiments, and can be particularly relevant to the understanding of the 
scalar sector. 

Recently, Fermilab experiment E791 reported on the indication of a light scalar
$K\pi$ resonance \cite{dkpipi}, $\kappa$, based on a full Dalitz plot analysis of the decay
$D^+\to K^-\pi^+\pi^+$. The measured Breit-Wigner mass and width of this state were found to be 
$797\pm 19\pm 43$ MeV/c$^2$ and $410\pm 43\pm 87$ MeV/c$^2$, respectively. 
Fermilab E791 has also observed evidence for the scalar $\sigma$ meson in
$D^+ \to \pi^- \pi^+ \pi^+$ decays \cite{d3pi} and measured $f_0$ masses and widths
in $D_s^+ \to \pi^- \pi^+ \pi^+$ decays \cite{ds3pi}.

Here, we briefly review the $D^+ \to K^- \pi^+ \pi^+$ methodology and the
results obtained. Moreover, we present new, preliminary studies 
providing further information on the phase variation of the scalar sector at low mass. 
We first present a study of the interference of $K^*(890)$ and underlying amplitudes 
as an attempt to measure the phase motion of the scalar low-mass $K\pi$ amplitude. 
A more extensive study comes afterwards, where we try a set 
of fit models inspired by the LASS parametrization for the S-wave $K\pi$ amplitude \cite{lass}. 
The main idea
is to check whether a unitary, isolated, S-wave $K\pi$ amplitude can fit the $D^+\to K^-\pi^+\pi^+$ Dalitz
plot, together with the other (higher-spin) intermediate states. We show that this approach is
not adequate for the data, and present similar parametrizations where final-state interactions are
included for the S-wave components, as is usually done in Dalitz plot analyses. In particular, 
we present a result offering further support for a rapidly-varying phase at low $K\pi$ mass, 
consistent with the previous indication of a light, broad $\kappa$ in $D^+\to K^-\pi^+\pi^+$.

\section{REVIEW OF THE $D^+\to K^-\pi^+\pi^+$ RESULTS}

\begin{figure}[t]
\begin{minipage}{7 cm}
\includegraphics[width=7cm]{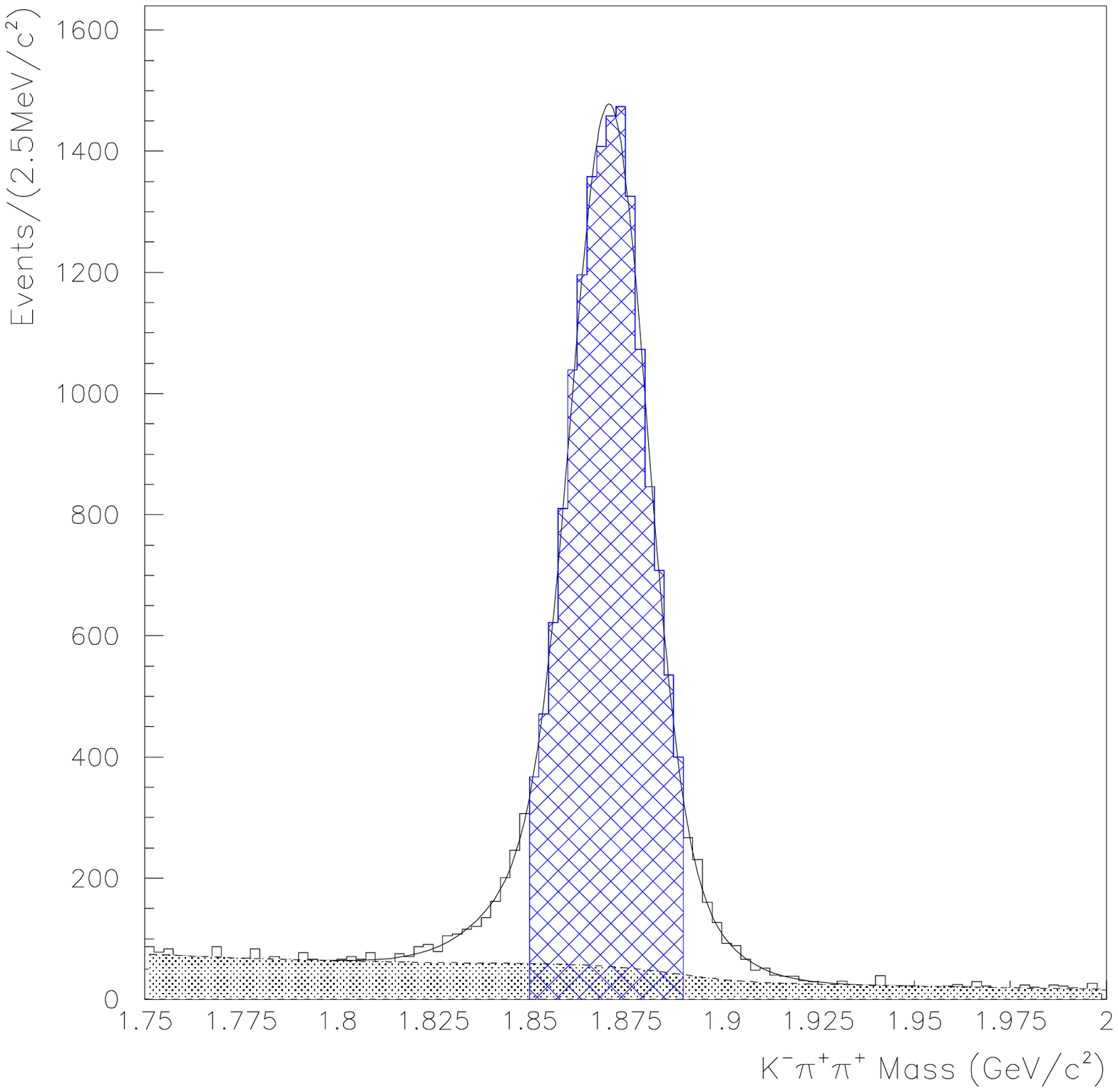}

\centerline{(a)}
\end{minipage}
\begin{minipage}{7 cm}
\includegraphics[width=7cm]{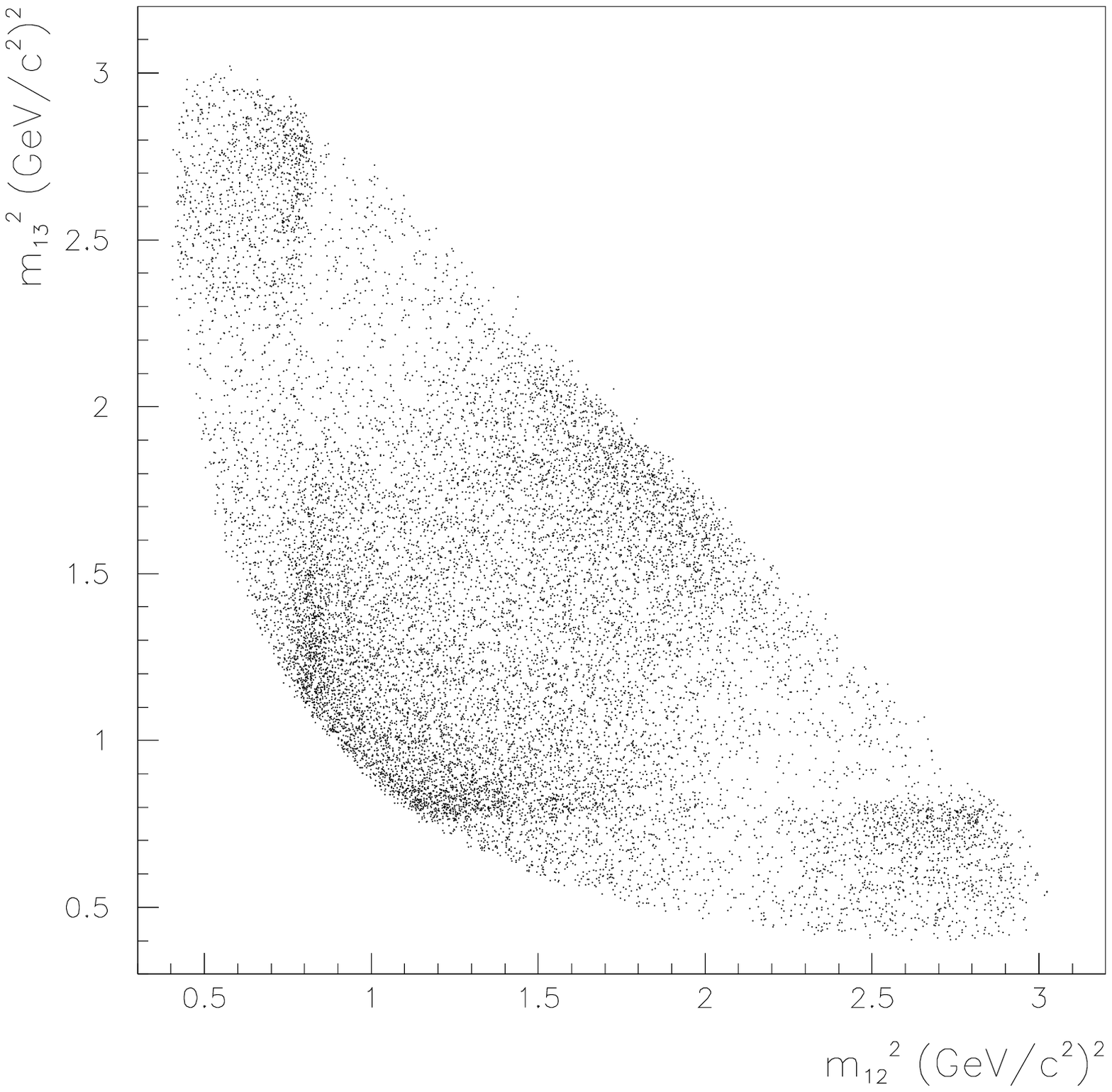}

\centerline{(b)}
\end{minipage}
\caption{(a) The $K\pi\pi$ invariant mass spectrum. The filled area is 
background; (b) Dalitz plot corresponding to the events in the crosshatched area of (a).
\label{kpipi} }
\end{figure}

From the original $2\times 10^{10}$ events collected by Fermilab E791 \cite{e791}, and after
reconstruction and selection criteria, we obtained the $K^-\pi^+\pi^+$ mass spectrum
shown in Fig.\ref{kpipi}(a). The sample of 15,090 events shown in the crosshatched area is 
used for the Dalitz plot analysis of $D^+\to K^-\pi^+\pi^+$, from which about 6\% is
due to background. The Dalitz plot of these events are shown in Fig.\ref{kpipi}(b), 
where the axes represent the two invariant mass squared combinations $s_{12}$ and $s_{13}$ 
(the kaon is labeled particle 1, the two pions are particles 2 and 3).

For the Dalitz plot analysis, an unbinned maximum-likelihood fit was performed 
with probability distribution funtions (PDF's) for both signal and background.
The signal PDF was written as the square of the total physical amplitude ${\cal A}$ and it is weighted
by the acceptance across the Dalitz plot and by 
the level of signal to background for each event.

Many $K^*$ resonances could contribute to the final state by forming quasi 2-body intermediate 
states $\bar K^*\pi^+$, as well as a possible non-resonant (NR) 3-body decay.
Due to the possibility of rescattering of these intermediate states 
from final state interactions (FSI), a general ansatz, widely use in Dalitz analyses of $D$ decays
to describe the total decay amplitude, is given by 
\begin{equation}
{\cal A} = a_{\rm NR} e^{i\delta_{\rm NR}}{\cal A}_{\rm NR} + \sum_{n=1}^N a_ne^{i\delta_n}
{\cal A}_n ~.
\end{equation} 
This equation uses Lorentz invariant amplitudes ${\cal A}_n$
to describe the individual resonant and NR processes, with relative strenghts given by
the magnitudes $a_n$ and strong phases $\delta_n$ to account for FSI. 
The NR amplitude ${\cal A}_{\rm NR}$ is usually assumed constant and set to 1. The resonant 
amplitudes are written as 
${\cal A}_n = \ BW_n~F_D^{(J)}~ F_n^{(J)}~ {\cal M}_n^{(J)} $
where $BW_n$ is the relativistic Breit-Wigner propagator, 
$ BW_n = \left[ m^2_0 - m^2 - im_0\Gamma(m) \right]^{-1}$, 
the quantities $F_D$ and  $F_R$ are the Blatt-Weisskopf damping factors
respectively for the $D$ and the $K\pi$ resonances and ${\cal M}_n^{(J)}$ describes 
the angular function according to the spin $J$ of the resonance. See \cite{dkpipi}. 
The amplitudes are generally functions of both Dalitz variables $s_{12}$ and $s_{13}$.
Here, with two identical pions in the final state, each amplitude is Bose symmetrized 
${\cal A}_n = {\cal A}_n[({\bf 12}){\bf3}] + {\cal A}_n[({\bf 13}){\bf 2}]$.

The first approach to fit the data includes the known $K^*$ resonances
plus a constant NR term. This is called Model A.
It finds contributions from the channels: NR, with a decay fraction
over 90\%, followed by $\bar K^*_0(1430)\pi^+$, 
$\bar K^*(892)\pi^+$, $\bar K^*(1680)\pi^+$ and $\bar K^*_2(1430)\pi^+$; 
these values are in accordance with previous results from Fermilab E691\cite{e691-kpipi} and 
Fermilab E687\cite{e687-kpipi}. See fit fractions
and relative phases in Table \ref{tablekappa}.
There is an important destructive interference pattern, since all fractions add up
to 140\%. Important disagreements appear between the fit to Model A and data, with
$\chi^2/\nu=2.7$ ($\nu$ being the number of degrees of freedom) in a binned version of 
the Dalitz plot. The discrepancies
are found mainly at low $K\pi$ mass squared (below 0.8 ${\rm GeV^2/c^4}$) and near
2.5 ${\rm GeV^2/c^4}$ \cite{dkpipi}. 
Thus, a model with 
the known $K\pi$ resonances, plus a constant NR amplitude, is not able to describe the 
$D^+\to K^-\pi^+\pi^+$ Dalitz plot satisfactorily.

A second model (Model B)  allows the mass and width of the scalar $K^*_0(1430)$ to
float, and includes Gaussian form-factors \cite{dkpipi}. The resulting fractions and phases are
similar to those of Model A (within errors) and the mass and width of $K^*_0(1430)$ are 
found to be consistent with PDG values. The fit improves but it is still unsatisfactory.

A third fit model, Model C, is constructed by the inclusion of an extra scalar state, 
with unconstrained mass and width. The mass and width of the $K^*_0(1430)$ are also  
free parameters, and Gaussian form-factors are used as in Model B. 
Using this model, the values obtained are  $797\pm 19\pm 43$ MeV/c$^2$ for the mass and 
$410\pm 43\pm 87$ MeV/c$^2$ for the width of the new scalar state, here referred to as the $\kappa$.
The values of mass and width obtained for the $K^*_0(1430)$ are respectively 
$1459\pm 7\pm 5$ MeV/c$^2$ and $175\pm 12\pm 12$ MeV/c$^2$, appearing heavier and
narrower than presented by the PDG. The fit fractions and phases from Models C are shown
in Table \ref{tablekappa}. Compared to the results of Model A (without $\kappa$), 
the NR mode drops from over 90\% to 13\%. The $\kappa\pi^+$ state 
is now dominant, with a decay fraction about 50\%. Moreover, the fit quality of Model 
C is substantially superior to that of Model A; the $\chi^2/\nu$ is now 0.73.

\begin{table}\centering
\caption{Results without $\kappa$ (Model A) and with $\kappa$ (Model C). }
\begin{tabular}{|c|c c|c c|} \hline 
 Decay & \multicolumn{2}{|c|}{Model A: No $\kappa$} 
 & \multicolumn{2}{|c|}{Model C: With $\kappa$ }  \\
 Mode  & Fraction (\%) & Phase & Fraction (\%) & Phase  \\  \hline
NR           & $90.9\pm 2.6$ & $0^\circ$ (fixed) & $13.0\pm 5.8\pm 4.4$ & $(-11\pm 14\pm 8)^\circ$  \\ 
$\kappa\pi^+$  & -- & -- & $47.8\pm 12.1\pm 5.3$ & $(187\pm 8\pm 18)^\circ$ \\ 
$\bar K^*(892)\pi^+$  & $13.8\pm 0.5$ & $(54\pm 2)^\circ$ & $12.3\pm 1.0\pm 0.9$ & $0^\circ$ (fixed)  \\ 
$\bar K^*_0(1430)\pi^+$ & $30.6\pm 1.6$ & $(54\pm 2)^\circ$ & $12.5\pm 1.4\pm 0.5$ & $(48\pm 7\pm 10)^\circ$ \\ 
$\bar K^*_2(1430)\pi^+$   & $0.4\pm 0.1$ & $(33\pm 8)^\circ$ & $0.5\pm 0.1\pm 0.2$ & $(-54\pm 8\pm 7)^\circ$\\ 
$\bar K^*(1680)\pi^+$    & $3.2\pm 0.3$ & $(66\pm 3)^\circ$ & $2.5\pm 0.7\pm 0.3$ & $(28\pm 13\pm 15)^\circ$\\ \hline 
\end{tabular}
\label{tablekappa}
\end{table}

From the results of the Dalitz plot analysis of $D^+\to K^-\pi^+\pi^+$, one concludes that
a conventional approach, including known $K\pi$ resonances and a constant NR term, cannot describe
the data. The presence of a broad, light scalar $K\pi$ state, $\kappa$, provides a very good description
of the data and becomes the major contribution to the decay. 
Many other possibilities were tried with no sucess, for instance a toy model for which the extra state is
represented by a Breit-Wigner without phase variation, vector and tensor hypotheses, 
different parametrizations for the NR channel, etc.


\section{Further Studies}

As described above, we obtain a very good description of the $D^+\to K^-\pi^+\pi^+$ data by the 
inclusion of the $\kappa$ in the decay amplitude. A scalar resonant phase behavior seems to be 
required at low mass. Nevertheless, it is desirable to investigate further, and independently of 
the Breit-Wigner hypothesis. 

Here, we present new studies in order to provide further information on the phase behavior
of the scalar amplitude at low mass. A first study refers to the asymmetry of the $K^*(890)$ and we show
that there is no sensivity to the $\kappa$ phase motion at the $K^*(890)$ mass in the crossed channel, 
although an indication of it can be seen at lower and higher mass. A second, more important study, 
refers to a comparison with the results of the S-wave $K\pi$ amplitude from LASS, 
composed of a non-resonant background (with effective range parametrization) and the 
$K^*_0(1430)$. The main results we obtain are the  inability to describe the data by 
imposing a $K\pi$ elastic unitarity constraint but, on the other hand,
the features of the phase motion obtained after this imposition is released.

\subsection{$K^*(890)$ Asymmetry Study}

A suggestion has been made to us \cite{ochsprivate} to see whether the phase motion of the $\kappa$
could be inferred through the interference of the $K^*(890)$ with the broad, underlying amplitude. 
From Fig.\ref{kpipi}(a) a strong asymmetry in the interference of the $K^*(890)$ is evident in the Dalitz plot.
This asymmetry is due to the interference with the dominant scalar component at low mass, i.e., the NR 
component in Model A (without $\kappa$) or the broad $\kappa$ in Model C. In the following, we will 
explore this possibility.

\begin{figure}
\includegraphics[width=9 cm]{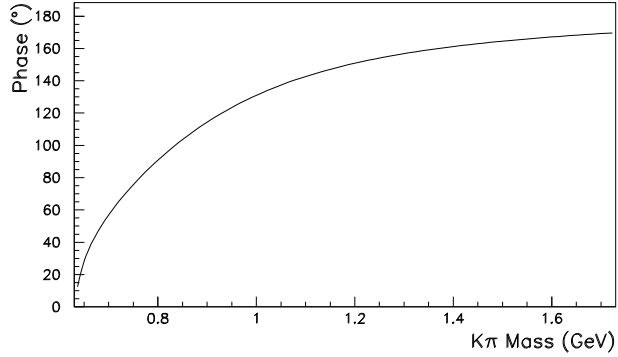}
\caption{The Breit-Wigner phase $\delta_{BW}$ for a scalar $K\pi$ resonance with mass 797 MeV/c$^2$ and
width 410 MeV/c$^2$}
\label{phaseBW}
\end{figure}

If one plots the phase of the $\kappa$ Breit-Wigner ($\delta_{BW}$) with central mass and 
width obtained by E791 (797 MeV/c$^2$ and 410 MeV/c$^2$, respectively) one sees that there is a sharp 
increasing of the phase from threshold up to about 1 GeV/c$^2$ (see Fig.\ref{phaseBW}). 
Above 1 GeV/c$^2$, the increase is slow. Now, we want to study the $\kappa$ phase motion in 
$m_{13}$ by inspection of the $\cos\theta_V$ distribution of $K^*(890)$ 
($\theta_V$ being the decay angle of the $K$ in the center-of-mass frame of $K^*(890)$ with 
respect to the $D$ direction) within 0.8 and 1.0 GeV/c$^2$ in the crossed channel, $m_{12}$. 
By observing the Dalitz Plot, we see that for $0.8<m_{12}<1.0$ GeV/c$^2$, the allowed kinematical 
region of the crossed channel is $m_{13}$ \lower.5ex\hbox{$\buildrel > \over \sim$} 1 GeV/c$^2$
\footnote{It is very important to call attention here to one feature of the
$D^+\to K^-\pi^+\pi^+$ decay. When looking at a low-mass region in one channel, let's say $s_{12}<0.7$ GeV/c$^2$$^2$ 
(just below $K^*(890)$ mass) we are at the same time observing a higher region of $s_{13}$,  
above 1.3 ${\rm GeV^2/c^4}$. This is a result of having two identical pions in the decay.
Thus, the presence of the $\kappa$ in this decay, especially its phase variation, 
would reflect itself at low and high $K\pi$ mass (respectively below 0.8 GeV/c$^2$ and above 1.5 GeV/c$^2$).
}. 
Thus, we can anticipate that the study of the $K^*(890)$ asymmetry will miss the main $\kappa$ 
phase variation in the crossed channel.

\begin{figure}[t]
\vspace{4.6 cm}
\begin{minipage}{6.7 cm}

\vspace{2. cm}
\hspace*{-1cm}\includegraphics[width=8 cm]{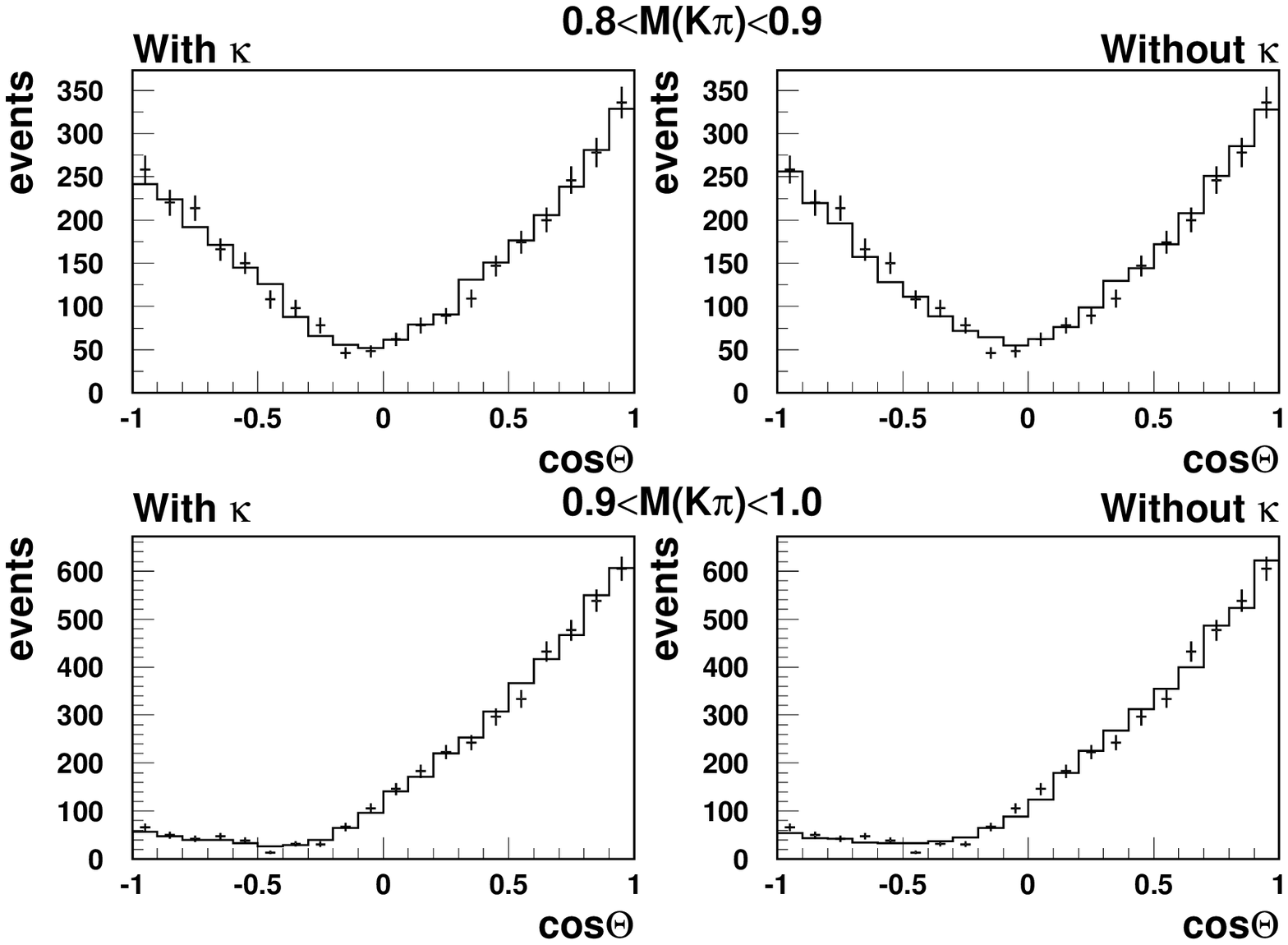}

\centerline{(a)}
\end{minipage}
\begin{minipage}{6.7 cm}
\includegraphics[width=8 cm]{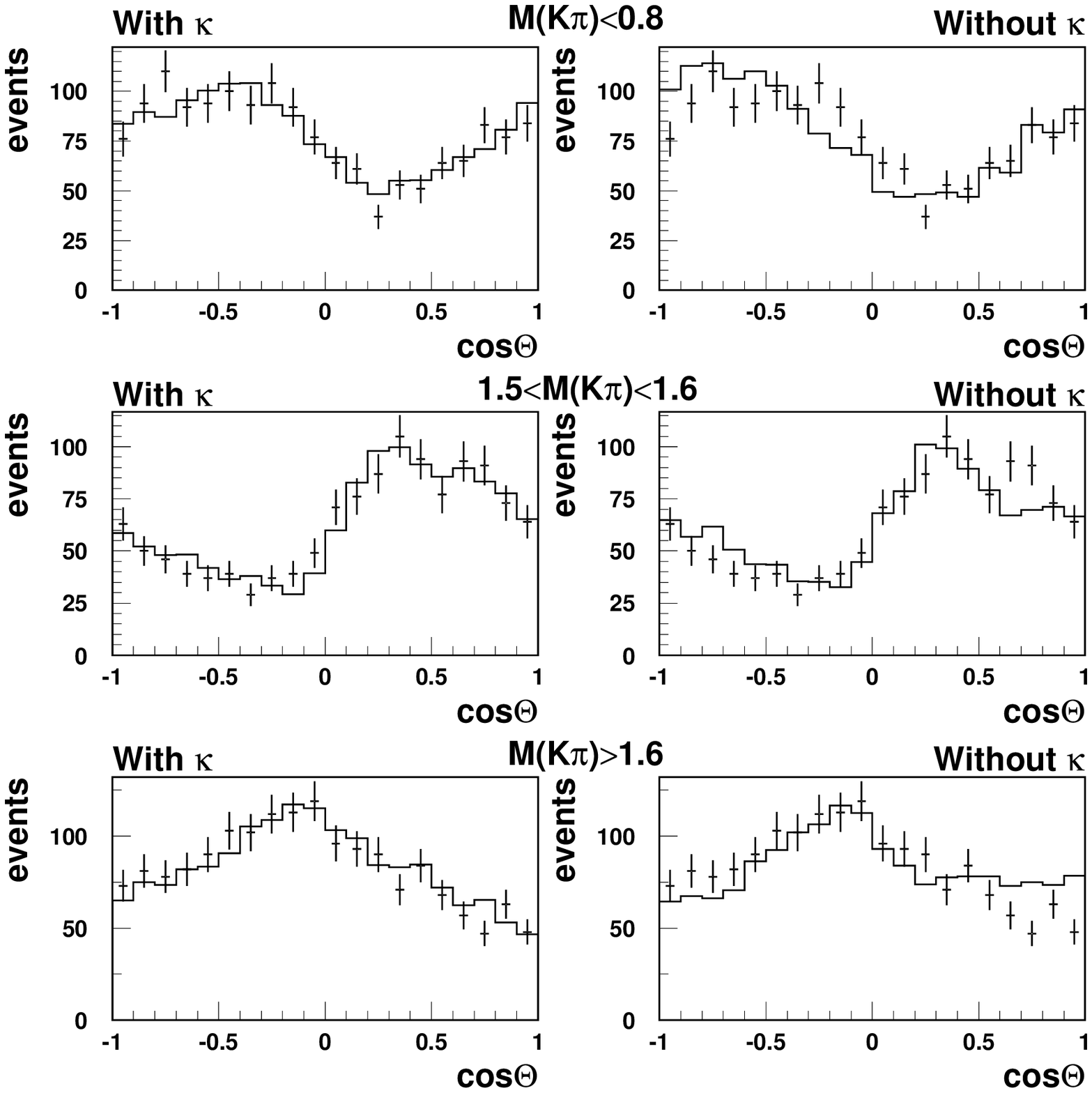}

\centerline{(b)}
\end{minipage}
\caption{ The $\cos\theta_V$ distribution: (a) within the $K^*(890)$ region (0.8--0.9 GeV/c$^2$);
(b) for low ($<0.8$ GeV/c$^2$) and high ($>1.5$ GeV/c$^2$) $K\pi$ mass 
\label{costheta} }
\end{figure}

From the data, this fact becomes evident. In Fig.\ref{costheta}, we plot the 
$\cos\theta_V$ distributions (a) within the $K^*(890)$ mass region and (b) for low 
and high $K\pi$ masses (outside the $K^*(890)$ region). For all plots, we compare the
data (points with error bars) to the result of our fit models with and without $\kappa$.
From Fig.\ref{costheta}(a) it is clear that it is not possible to distinguish between the
models. Nevertheless, for low and high $K\pi$ masses shown in Fig.\ref{costheta}(b) we
see that the model without $\kappa$ gives a poor description of the distributions, while
the model with $\kappa$ describes them adequately. 

\begin{figure}
\includegraphics[width=9.5 cm]{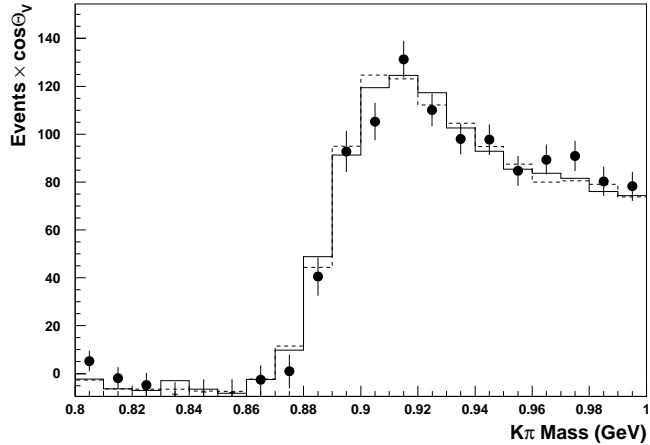}
\caption{The asymmetry distribution in $\cos\theta_V$ for $K\pi$ mass between 0.8 and 1.0 GeV/c$^2$. Error bars
represent the data, Model A is shown as the dotted histogram, and Model C as the solid-line histogram.}
  \label{costhetaweight}
\end{figure}

As a complementary view, the asymmetry distribution is obtained by plotting $\cos\theta_V$ 
weighted by the number of events in each
$K\pi$ mass bin between 0.8 and 1.0 GeV/c$^2$. For E791 data, the distribution is shown in 
Fig.\ref{costhetaweight} with error bars. In the same figure,  both models with (Model A) and without 
$\kappa$ (Model C) are compared to the data distribution. Both reproduce very well the asymmetry 
observed for the $K^*(890)$. Thus, using only this information
one would not claim the necessity of the $\kappa$ in the $D^+\to K^-\pi^+\pi^+$ decay. On the other
hand, the effects of the $\kappa$ can be seen below the $K^*(890)$ mass or further above it, as shown 
before.

It is worthwhile to comment on the asymmetry reported by Fermilab FOCUS \cite{focus} in their analysis of the
decay $D^+\to K^-\pi^+\mu^+\nu_\mu$, where the $K^*(890)$ was found to interfere with a small scalar
component with a phase of $45^\circ$. There, no FSI phases are expected, thus a comparison
with the phases of $D^+\to K^-\pi^+\pi^+$, as suggested by \cite{ochsprivate}, could be far 
from being direct (see more discussion in the next section).

\subsection{Comparisons with the LASS $K\pi$ Amplitude}

A set of studies was performed to compare the features of the S-wave $K\pi$
components in the $D^+\to K^-\pi^+\pi^+$ decays to the S-wave measured by  LASS
experiment \cite{lass} on elastic $K\pi$ scattering. It is interesting to discuss here the
differences between these two environments. 
We will be comparing production to a scattering process.
For $D$ decays, it is known that FSI can play an important role; explicitly 
there is the possibility of rescattering between quasi two-body states, $R_1\pi^+ \to R_2\pi^+$
(where $R_{1,2}$ are any of the $K\pi$ resonances). Thus the presence of the bachelor pion can have an influence
on the process, in this case not being isolated from $K\pi$. One consequence of this fact would be the loss
of unitarity in the $K\pi$ system. This is what we want to check.

The idea here is to try the same parametrization used by LASS to see whether the S-wave phase motion
observed by them could represent the $D^+\to K^-\pi^+\pi^+$ data. Their S-wave (up to the inelastic
threshold) had two terms: a non-resonant term, described through an effective-range form, and a relativistic
Breit-Wigner for $K^*_0(1430)$. No low-mass scalar resonance was observed. We will try this same
kind of approach in two ways. First, we impose the exact form they used, which implies the imposition 
of $K\pi$ elastic unitarity. Then, we release this restriction. 

The amplitude of the S-wave used by LASS, translated to a D decay 
(instead of scattering) needs a phase space correction given by
$m/p^*$ (mass of the resonance and decay momentum in the resonance frame, respectively). 
In the first fit, Fit 1, the relative strength and phase for NR and $K^*_0$ are 
fixed by imposing the unitarity constraint on the $K\pi$ system, so the total S-wave amplitude is
written as:
\begin{equation}
{\cal A}_S = {m_{12}\over p^*_{12}} \sin{\delta_B}\,e^{i\delta_B} + e^{2i\delta_B} 
{(m_0^2/p^{*0}_{12})\Gamma_0 \over m_0^2 - m_{12}^2 - i m_0\Gamma(m_{12})} ~+~(2\leftrightarrow 3)
\label{lassfix}
\end{equation}
with $\Gamma(m)= (m_0/m)(p^*/p^{*0})\Gamma_0$  and the NR term has the effective range form:
\begin{equation}
\cot{\delta_B} = {1\over a\,p^*} + {1\over 2} b\,p^*
\label{effrange}
\end{equation}
\begin{figure}
\includegraphics[width=8 cm]{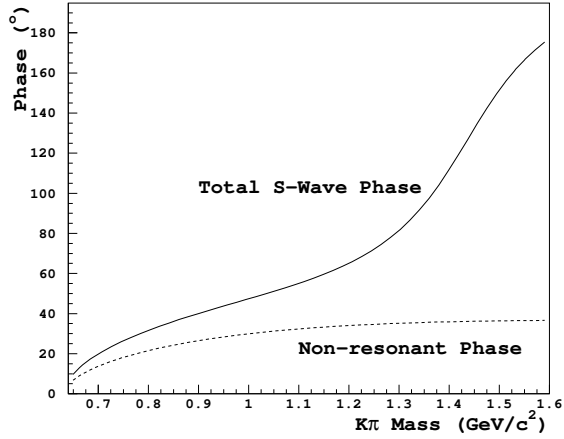}
\caption{The S-wave phase motion obtained by the LASS experiment (solid line). The behavior
of $\delta_{\rm NR}$ is shown through the dashed line.}
  \label{lassphase}
\end{figure}
The  values obtained by LASS in their fit were \cite{dunwoodie}:
\begin{eqnarray}
a = 1.95 \pm 0.09~{\rm c/GeV} &,& ~~~b = 1.76 \pm 0.36~{\rm c/GeV} \nonumber \\
M_{K^*_0(1430)} = 1435 \pm 5\,{\rm MeV/c^2} &,& ~~~\Gamma_{K^*_0(1430)} = 279 \pm 6\,{\rm MeV/c^2}
\label{lasspars}
\end{eqnarray}
From these values, we plot the S-wave phase motion they found from their data in Fig.\ref{lassphase}.

The total amplitude describing the $D^+\to K^-\pi^+\pi^+$ decay is written as a coherent sum of
the above S-wave term (Eq.\ref{lassfix}) and the other resonant states with spin $\ge 1$. 
The relative strenghts and phases ($a's$ and $\delta's$ below) are let free in the fit:
\begin{equation}
{\cal A}_{\rm TOT} = a_S e^{i\delta_S} {\cal A}_S + \sum_n^{J\ge 1} a_n e^{i\delta_n}
{\cal A}_n 
\end{equation}
When we fit the data with this model, we obtain a very bad representation. The $\chi^2/\nu$ 
is 10.6. This is evidence for the fact that 
the imposition of the $K\pi$ unitarity constraint for the S-wave does not work here: 
the S-wave $K\pi$ system is not isolated.

Let us now drop the constraint of elastic unitarity in the $K\pi$ system, and use a model closer
to what is the usual Dalitz plot approach, where the presence of FSI is included through 
extra strong phases for {\it all} contributing channels. As before,
the scalar components included are the NR (parametrized as effective range) and the 
$K^*_0(1430)$ but each of them now has an independent magnitude and phase.
Thus, the total decay amplitude for $D^+\to K^-\pi^+\pi^+$ is being written as:
\begin{eqnarray}
{\cal A}_{\rm TOT} &=& a_{\rm NR} e^{i\delta_{\rm NR}} 
 \frac{m^{}_{12}}{p^*_{12}} \sin{\delta_B}\, e^{i \delta_B}  \,+\, 
a_{K^*_0} e^{i\delta_{K^*_0}} \, BW_{K^*_0}(m_{12})  \,+ \,(2\leftrightarrow 3) \nonumber \\
&+& \sum_n^{J\ge 1} a_n \,e^{i\delta_n} \,{\cal A}_n .
\label{lass_nounit}
\end{eqnarray}
Using this approach, we try two fits: in Fit 2, we fix the parameters of the scalar
components to the values obtained by LASS, shown in Eq.\ref{lasspars}.
For Fit 3, these parameters ($a,~b,~M_{K^*_0(1430)},~\Gamma_{K^*_0(1430)}$) are allowed
to float. 

Our results for Fit 2 are much better than those for Fit 1. Nevertheless they are very much
like the results for Model A - the fit with a constant NR term, and without $\kappa$. 
The quality of Fit 2 is still not good ($\chi^2/\nu = 2.3$) and the NR term appears with an
enormous fraction of about 120\%.

Our results for Fit 3, on the other hand, are very interesting. We use the amplitude as in Eq.\ref{lass_nounit},
but releasing the parameters $a$ and $b$, and the mass and width of $K^*_0(1430)$. In this case, we obtain
a good description of the data. The values obtained for these parameters are:
\begin{eqnarray}
a = 4.58\pm 0.32 ~{\rm c/GeV} &,& ~~~b = -2.94 \pm 0.43 ~{\rm c/GeV} \nonumber \\
M_{K^*_0(1430)} = 1459 \pm 6 \,{\rm MeV/c^2} &,& ~~~\Gamma_{K^*_0(1430)} = 156 \pm 11 \,{\rm MeV/c^2}
\end{eqnarray}
Comparing to the best model, where $\kappa$ is included (Model C),
we observe: the fit qualities are comparable (the model with $\kappa$ is slightly better);
the mass and the width of the $K^*_0(1430)$ agree; the fraction of the non-resonant in Fit 3 ($66\pm 7$\%) 
is similar to the sum of the NR and $\kappa$ in Model C; results for the other states agree.

\begin{figure}
\includegraphics[width=8 cm]{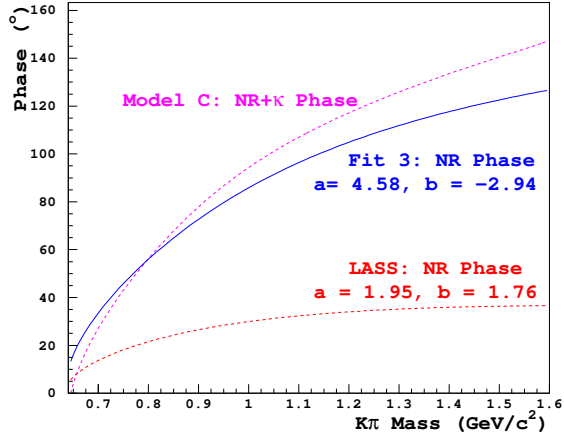}
\caption{NR phase in Fit 3 compared to Model C and to LASS}
  \label{phaseFit3}
\end{figure}

A very interesting feature of Fit 3 is the negative value for $b$. In this case, the
``effective-range'' would be indicating a resonant behavior. To better show this effect, we plot in
Fig.\ref{phaseFit3} the $\delta_B$ phase motion of the ``NR'' decay in Fit 3 for the fitted values 
of $a$ and $b$. Note that the phase increases very rapidly from threshold. 
This effect is similar to the effect we found for $\kappa$. 
Here in Fit 3, the ``effective-range'' phase would be replacing that for
a constant NR and the $\kappa$ in Model C (also shown in the figure);
the two curves are very similar, even more if one realizes that one is mainly the result of a Breit-Wigner 
shape for the $\kappa$ and the other comes from what would be an effective-range form (but turned 
out to have an impressive phase variation). For comparison, the curve of $\delta_B$ with the parameters 
obtained by LASS are also shown: an almost constant phase for the NR was observed.

We see that the scalar sector at low $K\pi$ mass in $D^+\to K^-\pi^+\pi^+$ needs a rapidly-varying
phase, consistent with the previous indication of the $\kappa$ state from \cite{dkpipi}.

\section{Conclusions}
Using the Fermilab E791 data sample of $D^+\to K^-\pi^+\pi^+$ , we present new, preliminary studies 
offering further information concerning the phase motion behavior of the scalar amplitude at low
$K\pi$ mass. In particular, we showed that the S-wave phase motion as measured by LASS in a 
scattering process is not able to describe the E791 decay data. This result shows the significance of
final-state interactions in $D$ decays preventing a direct comparison of the phases in scattering and decay processes. 
Furthermore, an independent parametrization (without using a Breit-Wigner) provided a phase behavior 
consistent to what we found previously for the scalar low-mass $K\pi$ amplitude as a $\kappa$ meson.


\bibliographystyle{aipproc}   

\end{document}